\documentclass[journal]{IEEEtran}
\usepackage{color}
\usepackage{cite}
\usepackage{amssymb,amsfonts,url}
\usepackage{algorithmic}
\usepackage{textcomp}
\usepackage[rgb]{xcolor}
\usepackage{float}
\usepackage{placeins}
\usepackage{psfrag}
\newcommand{\stz}{\rule{0mm}{2.3ex}}

\usepackage{amsmath,graphicx,bm}
\usepackage{color,soul,cite}
\usepackage{MnSymbol}
\usepackage{multirow}
\usepackage{makecell}
\usepackage{xcolor,colortbl,hhline}
\usepackage[super]{nth}
\usepackage{subcaption}
\usepackage[colorlinks=true,linkcolor=black,anchorcolor=black,citecolor=black,urlcolor=blue]{hyperref}
\usepackage{setspace}
\usepackage{dblfloatfix,booktabs,caption}

\ifCLASSINFOpdf
\else
\fi
\begin{document}
\title{Deep Noise Suppression\\ Maximizing Non-Differentiable PESQ\\ Mediated by a Non-Intrusive PESQNet}

\author{Ziyi Xu,
        Maximilian Strake
        and Tim Fingscheidt,~\IEEEmembership{Senior Member,~IEEE}% <-this % stops a space
\thanks{Z. Xu, M. Strake, and T. Fingscheidt are with Institute for Communications Technology, Technische Universit{\"a}t Braunschweig, 38106 Braunschweig, Germany (e-mails:$\left \{ \text{ziyi.xu, m.strake, t.fingscheidt} \right \}$@tu-bs.de)}}


\markboth{}%
{Shell \MakeLowercase{\textit{et al.}}: Bare Demo of IEEEtran.cls for IEEE Journals}


\maketitle

\begin{abstract}
Speech enhancement employing deep neural networks (DNNs) for denoising are called deep noise suppression (DNS). During training, DNS methods are typically trained with mean squared error (MSE) type loss functions, which do not guarantee good perceptual quality. Perceptual evaluation of speech quality (PESQ) is a widely used metric for evaluating speech quality. However, the original PESQ algorithm is non-differentiable, and therefore cannot directly be used as optimization criterion for gradient-based learning. In this work, we propose an end-to-end non-intrusive \texttt{\textbf{PESQNet}} DNN to estimate the PESQ scores of the enhanced speech signal. Thus, by providing a reference-free perceptual loss, it serves as a mediator towards the DNS training, allowing to maximize the PESQ score of the enhanced speech signal. We illustrate the potential of our proposed \texttt{\textbf{PESQNet}}-mediated training on the basis of an already strong baseline DNS. As further novelty, we propose to train the DNS and the \texttt{\textbf{PESQNet}} alternatingly to keep the \texttt{\textbf{PESQNet}} up-to-date and perform well specifically for the DNS under training. Our proposed method is compared to the same DNS trained with MSE-based loss for joint denoising and dereverberation, and the Interspeech 2021 DNS Challenge baseline. Detailed analysis shows that the \texttt{\textbf{PESQNet}} mediation can further increase the DNS performance by about 0.1 PESQ points on synthetic test data and by 0.03 DNSMOS points on real test data, compared to training with the MSE-based loss. Our proposed method also outperforms the Challenge baseline by 0.2 PESQ points on synthetic test data and 0.1 DNSMOS points on real test data.
\end{abstract}


\IEEEpeerreviewmaketitle

\section{Introduction}
%%%%%%%%%%%%%%%%%%%%%%%%%%%%%%%%%%%%%%%
Speech enhancement aims at improving the intelligibility and perceived quality of a speech signal degraded by interferences, which can include both additive noise and reverberation. This task becomes particularly challenging when only a single-channel microphone mixture signal is available. However, single-channel speech enhancement is important in real-world applications, including telephony, hearing aids, and robust speech recognition, and thus has attracted a lot of research attention. The classical solution for this task is to estimate a time-frequency (T-F) domain mask, or, more specifically, a spectral weighting rule \cite{Ephraim1984,Ephraim1985,Scalart1996,Lotter2005,Cohen2005a,Gerkmann2008b,fodor2011speech,fodor2012mmse,samy_SNR}, which however, requires the estimation of the noise power, the \textit{a priori} signal-to-noise ratio (SNR), and sometimes also the {\it a posteriori} SNR.

The rise of data-driven approaches (e.g., deep learning) has facilitated the (direct) estimation of T-F-domain masks, or even the desired clean speech spectrum without any intermediate steps, which provides state-of-the-art performance, even in the presence of non-stationary noise \cite{fingscheidt2006data,erkelens2006general,Fingscheidt2008,wang2014training,weninger2014discriminatively,wang2015deep,williamson2016complex,park2017fully,zhao2018convolutionalrecurrent,wang2018supervised,tan2019complex,strake2019separated,strake2020fully,strake2020DNS,xu2021inter}.  Fingscheidt and Erkelens pioneered the data-driven approach for ideal mask-based speech enhancement in \cite{fingscheidt2006data,erkelens2006general,Fingscheidt2008}, which reduces the speech distortion while retaining a high noise attenuation. In recent years, deep neural networks (DNNs) have shown great success in speech enhancement. These methods have pushed the performance limits even further, due to the powerful modeling capability of the DNNs, and are subsumed under the term deep noise suppression (DNS). Wang et al.\ \cite{wang2014training,wang2018supervised} illustrate the benefits obtained by estimating an ideal ratio mask for supervised speech enhancement with DNNs. Williamson et al.\ \cite{williamson2016complex} propose the DNN-based estimation of a complex ratio mask from the single-channel mixture, to enhance both the amplitude spectrogram and also the phase of the noisy speech signal. Compared to feed-forward DNNs, convolutional neural networks (CNNs) are structurally well-suited to preserve the harmonic structures of the speech spectrum in denoising tasks, and have been successfully applied in \cite{park2017fully,zhao2018convolutionalrecurrent,tan2019complex,strake2019separated,strake2020fully}. Furthermore, the speech denoising task benefits significantly from networks which can model long-term temporal dependencies, as shown in recent studies integrating long short-term memory (LSTM) into CNNs \cite{zhao2018convolutionalrecurrent,tan2019complex,strake2019separated,strake2020fully}. A fully connected LSTM is integrated into the bottleneck of a convolutional encoder-decoder (CED) structure in \cite{tan2019complex}. However, the fully-connected LSTM usually contains a large amount of trainable parameters, which may weaken the generalization capability of the trained network. To mitigate this problem, Strake et al.\ \cite{strake2020fully} proposed to insert a convolutional LSTM (ConvLSTM) layer into the bottleneck of the CED structure instead of the fully-connected LSTM layer, which inherits the weight-sharing property of the CNN, thereby significantly decreasing the amount of trainable parameters. This network topology is dubbed as fully convolutional recurrent neural network ({\tt FCRN}), and was successfully employed in the \mbox{Interspeech} 2020 Deep Noise Suppression (DNS) Challenge \cite{strake2020DNS} for joint dereverberation and denoising, securing \nth{2} rank in the non-realtime track with a realtime model.

Besides looking for powerful DNS architectures, adopting a better loss function to train the DNS models is another important research direction to further improve the performance. During the training process, most of the DNS architectures are trained with some mean squared error (MSE) loss, which is minimized during the training process. However, optimization of the MSE loss during training does not guarantee good human perceptual quality of the enhanced speech signal (unless the MSE is zero), which leads to limited performance \cite{liu2017perceptually,kolbcek2018monaural,zhang2018training,martin2018deep,fu2018end,zhao2019perceptual,fu2019learning}. This effect is even more evident in low SNR conditions, where it can lead to a highly distorted speech component and very unnatural sounding residual noise. A perceptually-weighted loss function considering human psychoacoustics, such as the absolute hearing threshold and the masking effect, has recently been proposed in \cite{liu2017perceptually,zhao2019perceptual}. Zhao et al.\ \cite{zhao2019perceptual} employed the perceptual weighting filters from code-excited linear predictive (CELP) speech coding to weight the MSE between the network output and the corresponding target to emphasize perceptually important T-F regions. 

Perceptual evaluation of speech quality (ITU-T P.862.2 PESQ) \cite{ITUT_pesq_wb_corri} and short-time objective intelligibility (STOI) \cite{taal2010short} are two widely-used metrics for evaluating speech quality and intelligibility, respectively. Among these two metrics, PESQ has been designed to predict absolute category rating (ACR) listener scores in speech quality assessment. Thus, another approach would be to adapt perceptually related metrics, e.g., PESQ and STOI, as loss functions, which could be used to optimize for speech quality and speech intelligibility, respectively. However, both the original STOI and PESQ are non-differentiable functions, which cannot directly be used as an optimization criterion for gradient-based deep learning. Kolbcek et al.\ proposed a differentiable STOI loss formulation adapted from its original, and employed it in training a network for an utterance-based speech enhancement \cite{kolbcek2018monaural}. However, their experimental results reveal that the network trained with the proposed STOI loss does not significantly improve on the STOI score compared to the one trained with the conventional MSE loss. Compared to STOI, the original PESQ formulation is significantly more complex. Zhang et al.\ proposed a directional sampling method to approximate gradient descent employing the original PESQ formulation \cite{zhang2018training}. {M}art{\'\i}n-{D}o{\~n}as et al.\ proposed an approximated PESQ formulation, which is differentiable, and combined it with the MSE as the final optimization criterion \cite{martin2018deep}. This proposed PESQ loss gains in speech perceptual quality compared to the conventional MSE loss, however, does not outperform the simple perceptual weighting filter loss proposed by Zhao et al.\ \cite{zhao2019perceptual}. Thus, obtaining a better differentiable approximation of the original PESQ formulation is still a research challenge. 

To mitigate this problem, Fu et al.\ \cite{fu2019learning} trained an end-to-end deep learning model to approximate the PESQ function (so-called {\tt Quality-Net}) without having to know any computational details of the original PESQ formulation. Afterwards, the trained {\tt Quality-Net} is fixed and concatenated to the output of the DNS model to estimate the PESQ scores of the enhanced speech. Thus, it serves as a differentiable PESQ loss for the training of the DNS model, aiming at maximizing the PESQ of the output enhanced speech signal. It is important to note that as the original PESQ \cite{ITUT_pesq_wb_corri}, \cite{zhang2018training,martin2018deep}, and \cite{fu2019learning} require a clean reference signal for the loss calculation ({\it intrusive} quality model) and therefore do not allow to use real data in training. Furthermore, as reported by the authors of \cite{fu2019learning}, the gradient from their fixed {\tt Quality-Net} can guide the DNS model in increasing the PESQ scores only in the first few training iterations (minibatches). The reason could be that the fixed {\tt Quality-Net} has not seen the enhanced speech signal generated by the updated DNS model. Therefore, the gradient obtained from the fixed {\tt Quality-Net} is problematic after training for several minibatches, leading to the phenomenon that the {\tt Quality-Net} is fooled by the updated DNS model (estimated PESQ scores increase while true PESQ scores decrease).
\begin{figure}[t!]
	\centering
	\includegraphics[width=0.49\textwidth]{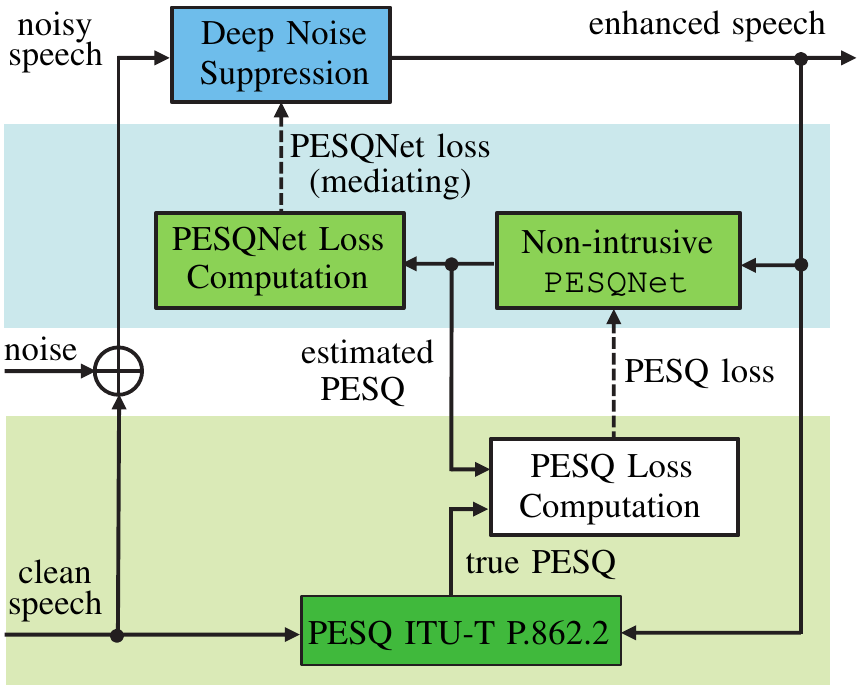}
	\caption{{\bf Proposed DNS model training mediated by the} \texttt{\textbf{PESQNet}}.}
	\label{system1}
\end{figure}

In this work, as shown in Fig.\,\ref{system1}, we propose an end-to-end {\it non-intrusive} {\tt PESQNet} modeling the PESQ function, and subsequently employ it as mediator towards the training of a DNS model. Compared to the {\tt Quality-Net} used in \cite{fu2019learning}, {\it our proposed {\tt PESQNet} can estimate the PESQ score of the enhanced signals without knowing the corresponding clean speech} (just like human raters in ACR listening tests). Thus, it can serve to provide a reference-free perceptual loss. {\it This offers the potential to train a DNS model employing real recorded data, where only the noisy speech mixture is available}, a specific problem that has been first addressed in \cite{xu2021inter}, however, still with insignificant improvements. Inspired from the alternating training schemes used in adversarial trainings \cite{pascual2017segan,pandey2018adversarial,meng2018cycle}, one important novelty of this work is solving the problems addressed in \cite{fu2019learning} by training the DNS model and the {\tt PESQNet} in a successful alternating protocol on epoch level. Thus, the {\tt PESQNet} can always {\it adapt to the current updated DNS model} and can serve as mediator between the DNS model training and the PESQ metrics by providing gradients, which effectively maximizes the original ITUT-T P.862.2 PESQ score of the enhanced speech signal. For the DNS model training, a fixed {\tt PESQNet} is employed to estimate the PESQ scores of the enhanced speech signal from the DNS model. Subsequently, a loss calculated on the estimated PESQ scores, called PESQNet loss, is used to control the DNS model training. Afterwards, the DNS model is fixed, and the so-called PESQ loss is calculated to reflect the difference between the estimated PESQ score from {\tt PESQNet} and its ground truth measured by the original PESQ function \cite{ITUT_pesq_wb_corri}. Thus, the {\tt PESQNet} training keeps up with the DNS model in any learning step.

We adopted the high-ranked {\tt FCRN} proposed by Strake et al.\ \cite{strake2020DNS} as our DNS model. The proposed {\tt PESQNet} model is adapted from a speech emotion recognition (SER) model proposed in \cite{Meyer2021}, which can provide state-of-the-art emotion recognition performance. An SER task is structurally similar to speech quality prediction, as it also assigns one label to an entire input speech utterance. Furthermore, the adapted SER model has been shown to be robust w.r.t.\ noise as well as channel and language variation, which is supposed to be important for our non-intrusive {\tt PESQNet}. Our proposed {\tt PESQNet} is used as mediator towards a fine-tuning on a pre-trained DNS model to further increase the perceptual quality. In the pre-training stage of our work, we consider both denoising and dereverberation by employing a joint MSE-based loss function proposed in \cite{strake2020DNS}, which has shown success in the Interspeech 2020 DNS Challenge \cite{reddy2020interspeech}. Note that our proposed learning strategy could potentially provide benefit to any given DNS model and to any given pre-training loss function.

The rest of the article is structured as follows: In Section\,\uppercase\expandafter{\romannumeral 2} we introduce the signal model and our mathematical notations. We describe our employed {\tt FCRN} and {\tt PESQNet} in Section\,\uppercase\expandafter{\romannumeral 3}. Then the details of mediating the {\tt FCRN} training with the proposed {\tt PESQNet} are explained. Next, we present the experimental setup including the database, training protocols, baselines, and employed quality metrics in Section\,\uppercase\expandafter{\romannumeral 4}. The results and discussion are given in Section\,\uppercase\expandafter{\romannumeral 5} and our work is concluded in Section\,\uppercase\expandafter{\romannumeral 6}.
%%%%%%%%%%%%%%%%%%%%%%%%%%%%%%%%%%%%%%%%
\section{Signal model and Notations}
We assume the microphone mixture $y(n)$ to be composed of the clean speech signal $s(n)$ reverberated by the room impulse response (RIR) $h(n)$, and disturbed by an additive noise $d(n)$ as
\begin{equation} \label{micro_mixture}
y(n)=s(n)*h(n)+d(n)= s^\text{rev}(n)+d(n),
\end{equation}
with $s^\text{rev}(n)$ and $n$ being the reverberated clean speech component and the discrete-time sample index, respectively, and $*$ denoting the convolution operation. Since we perform a spectrum enhancement, we transfer all the signals to the discrete Fourier transform (DFT) domain:
\begin{equation} \label{micro_fft}
Y_\ell(k)=S^\text{rev}_\ell(k)+D_\ell(k),
\end{equation}
with frame index $\ell$ and frequency bin index $k\!\in\!\mathcal{K}\!=\!\left \{0,1,\ldots,K\!-\!1\right \}$, and $K$ being the DFT size. This procedure is also often known as short-time Fourier transform (STFT), and successive STFT frames overlap in time. In this work, we adopt the {\tt FCRN} proposed by Strake et al.\ \cite{strake2020fully} as our DNS model, in which a magnitude-bounded complex mask $M_\ell\left(k \right )\in\mathbb{C}$ is estimated, with $\left|M_\ell\left(k \right )\right|\in\left [ 0,1 \right ]$ to enhance the noisy speech spectrum (see Fig.\,\ref{fig:CNN_topology}):
\begin{equation} \label{clean_speech_est}
\hat{S}_\ell\left (k \right )=M_\ell(k)\cdot Y_\ell(k).
\end{equation}
Finally, the enhanced speech spectrum $\hat{S}_\ell\left (k \right )$ is subject to an inverse DFT (IDFT), followed by overlap add (OLA) to reconstruct the time-domain enhanced speech signal $\hat{s}(n)$.
%%%%%%%%%%%%%%%%%%%%%%%%%%%%%%%%%%%%%%%%%
\section{Models and Novel Training Loss/ Protocol}
%%%%%%%%%%%%%%%%%%%%%%%%%%%%%%%%%%%%%%%%%%%%%%%%%%%%%%%%%%%%%%%%%%%%%%%%%%%%%%%%%%%%%%%
\subsection{DNS Model}
\begin{figure}[t!]
	%\vspace*{1.5mm}
	\centering
	\includegraphics[width=0.49\textwidth]{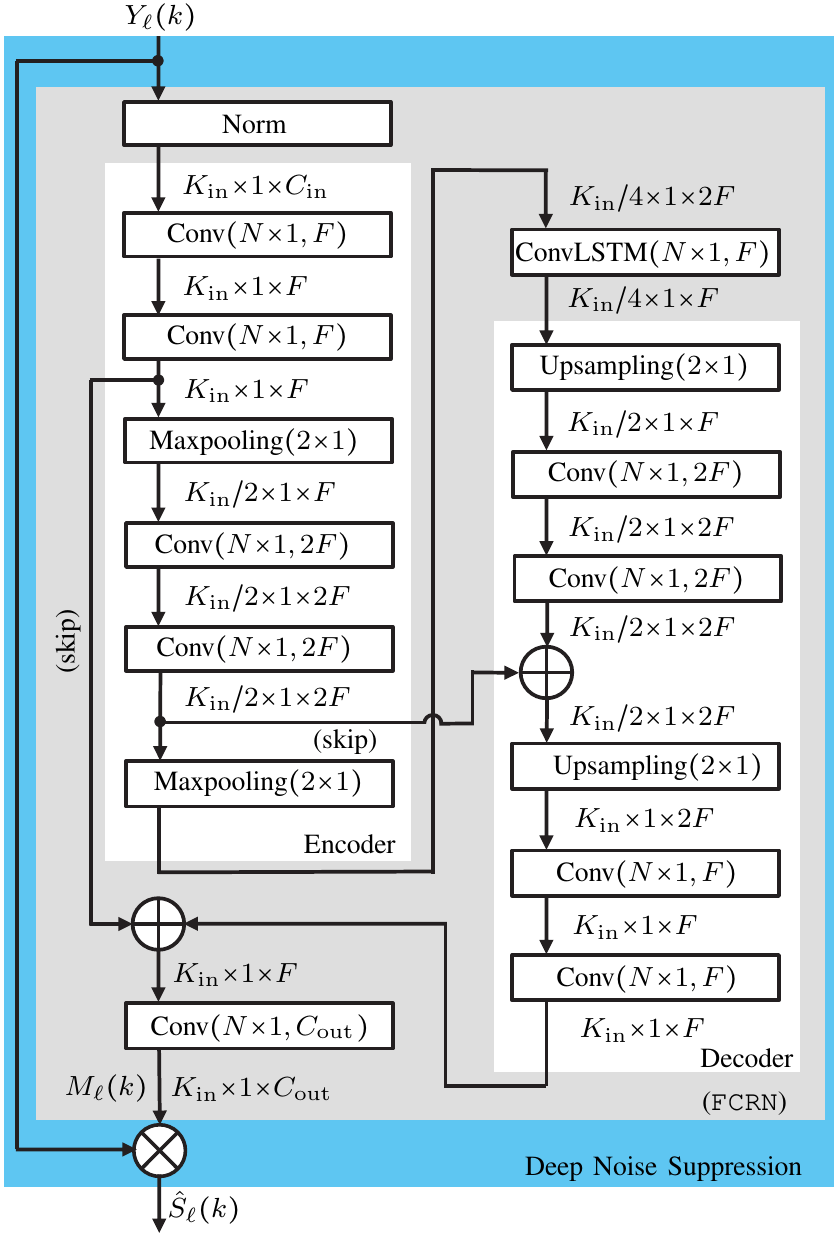}
	\caption{{\bf Employed DNS model} as used in Figs.\,\ref{system1}, \ref{system2}, \ref{system6}, and \ref{system3}.}
	\label{fig:CNN_topology}
\end{figure}

The DNS model employed in this work is shown in Fig.\,\ref{fig:CNN_topology}, with the {\tt FCRN} adopted from \cite{strake2020fully}. The input of the DNS is the noisy speech spectrum $Y_\ell(k)$. The Norm box represents a zero-mean and unit-variance normalization based on the statistics collected on the training dataset. The dimensions of the input and the output feature maps for each layer in the {\tt FCRN} are depicted as {\it number of features} $\times$ {\it number of time frames} $\times$ {\it number of feature maps}, with $K_{\rm in}$ representing the number of input and output frequency bins, while $C_{\rm in}$ and $C_{\rm out}$ denote the number of input and output channels, respectively.

The convolutional layers are represented by the \mbox{Conv$(N\times 1, f)$} operations, with $f\!=\!F$ or $f\!=\!2F$ being the number of filter kernels in each layer, and $(N\times 1)$ representing the kernel size, emphasizing that the convolutions are only performed along the feature axis. The maxpooling and upsampling layers have a kernel size of $(2\times 1)$. The stride of the maxpooling layers is set to $2$. The employed {\tt FCRN} contains a convolutional encoder-decoder (CED) structure. A fully convolutional LSTM layer denoted as ConvLSTM$(N\times 1, F)$ is integrated in the bottleneck of the CED structure as shown in Fig.\,\ref{fig:CNN_topology}. Furthermore, all possible forward residual skip connections are added wherever layers have matched dimensions.
\begin{figure}[t!]
	\centering
	\includegraphics[width=0.49\textwidth]{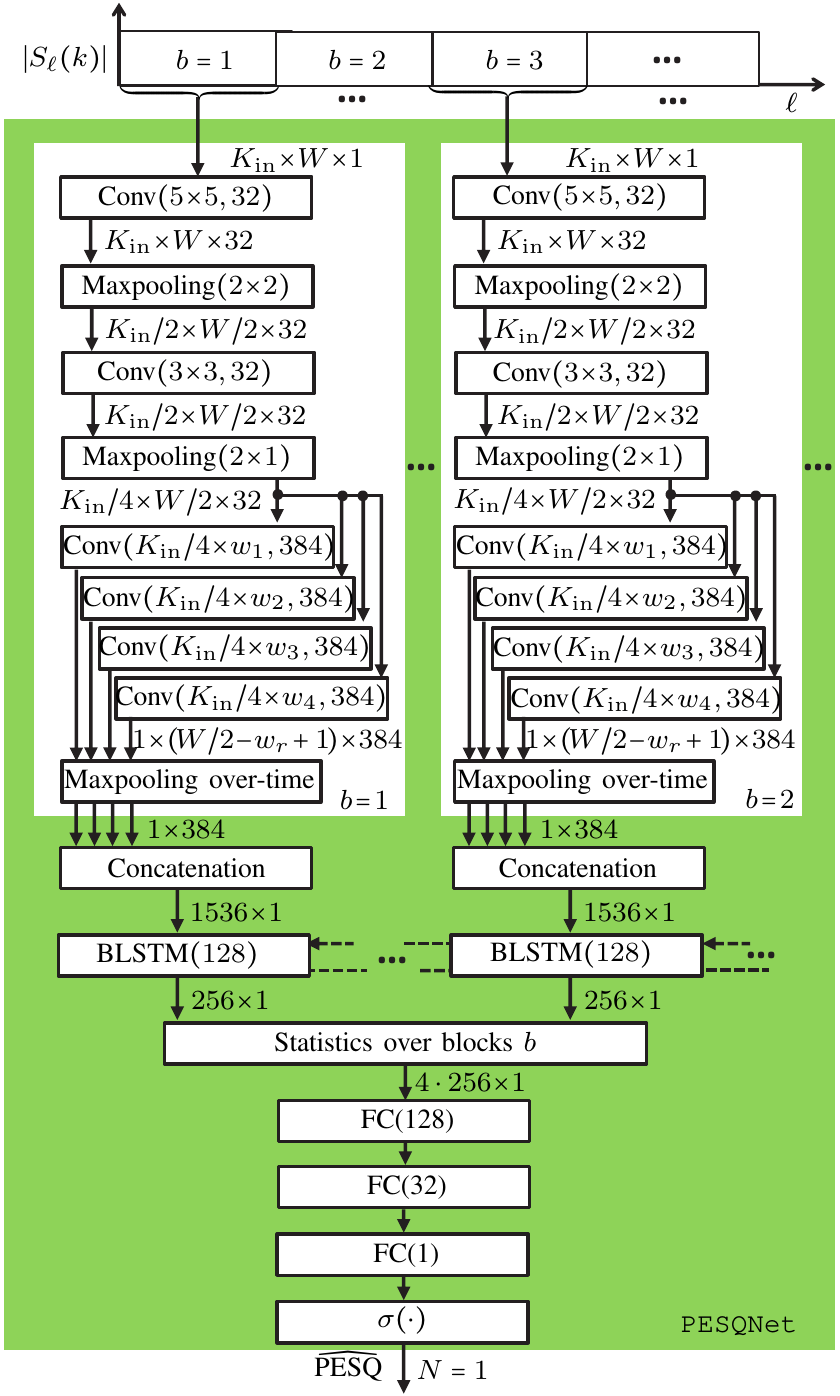}
	\caption{{\bf Employed} \texttt{\textbf{PESQNet}} as used in Figs.\,\ref{system1}, \ref{system6}, and \ref{system3}.}
	\label{fig:PESQNet}
\end{figure}
%%%%%%%%%%%%%%%%%%%%%%%%%%%%%%%%%%%%%%%%%%%%%%%%%%%%%%%%%%%%%%%%%%%%%%%%%%%%%%%%%%%%%%%%%
\subsection{PESQNet}
In this work, we propose an end-to-end {\it non-intrusive} {\tt PESQNet}, modeling ITU-T P862.2 PESQ. It estimates the PESQ score of an enhanced speech utterance in the DFT domain, and is subsequently employed to control the training of a DNS model. The employed {\tt PESQNet} shall deliver a single value (label) for an entire utterance, just as speech emotion recognition (SER) does. Furthermore, models used for SER are always non-intrusive, which is suitable for our reference-free PESQ estimation. Accordingly, for our {\tt PESQNet}, we built upon the SER model topology proposed in \cite{Meyer2021}, which shows state-of-the-art performance in emotion recognition. It is depicted in Fig.\,\ref{fig:PESQNet}.

The dimensions of the input and the output feature maps for each layer are depicted as {\it number of features} $\times$ {\it number of time frames} $\times$ {\it number of feature maps (if applicable)}. The input of the proposed {\tt PESQNet} is the amplitude spectrogram $\left|S_\ell(k)\right|$ (or: $\left|\hat{S}_\ell(k)\right|$), $\ell\!\in\!\mathcal{L}\!=\!\left \{1,2,\ldots,L\right \}$, of an entire utterance with $L$ frames, which is then grouped to several feature matrices (blocks indexed with $b\in\mathcal{B}=\left\{1, 2 \ldots, B\right\}$). Each feature matrix (block) has the same dimensions of $K_{\rm in}\!\times\! W\! \times\! 1$, with $K_{\rm in}$ and $W$ being the numbers of input frequency bins and time frames per block, respectively. The blocks are processed in parallel by identical subnetworks with the following structure. A CNN encoder is employed to extract quality-related features from the input feature matrices. The convolutional layers are represented by the Conv$(h\times w, f)$ operations, again with $f$ representing the number of filter kernels in each layer, and $(h\times w)$ representing the kernel size. Then, the extracted features are processed by multi-width convolutional kernels with kernel widths $w_1$, $w_2$, $w_3$, and $w_4$. The max-pooling-over-time layer and the subsequent concatenation deliver a feature map with a fixed dimension to the bidirectional LSTM (BLSTM) layer, which is used to model temporal dependencies. Afterwards, four statistics (average, standard deviation, minimum, and maximum) over blocks $b$ are applied to the BLSTM outputs before they are processed by the fully-connected (FC) layers denoted as FC$(N)$, with $N$ being the number of output nodes. The output layer has a single node with a gate function $\sigma(x)=3.6\cdot\text{sigmoid}(x)+1.04$ to limit the range of the estimated PESQ score between $1.04$ and $4.64$, as it is determined by ITU-T P.862.2 \cite{ITUT_pesq_wb_corri}.

The estimated PESQ score $\widehat{\text{PESQ}}_u$ of the utterance indexed with $u$ obtained from the {\tt PESQNet} should be as close as possible to its ground truth $\text{PESQ}_u$ measured by ITU-T P.862.2 \cite{ITUT_pesq_wb_corri}. Thus, the loss function for {\tt PESQNet} training is defined as (``PESQ loss"):
\begin{equation} \label{PESQNet}
J^\text{PESQ}_u\!=\left(\widehat{\text{PESQ}}_u-\text{PESQ}_u\right)^2.
\end{equation}
%%%%%%%%%%%%%%%%%%%%%%%%%%%%%%%%%%%%%%%%%%%%%%%%%%%%%%%%%%%%%%%%%%%%%%%%%%%%%%%%%%%%%%%%
\subsection{Novel PESQNet Loss Mediating Towards DNS Training}
Our proposed {\tt PESQNet} is utilized to control a fine-tuning on a pre-trained DNS model to further increase the perceptual quality. 
\subsubsection{DNS Pre-Training}
First, however, in the pre-training stage of our work, we consider a joint denoising and dereverberation by employing an MSE-based loss function as proposed in \cite{strake2020DNS}, which has shown good success in the Interspeech 2020 DNS Challenge \cite{reddy2020interspeech}. This joint loss function consists of two MSE-type loss terms. The first loss term is an utterance-wise loss aiming at joint dereverberation and denoising by utilizing the clean speech spectrum $S_\ell(k)$ as target, and is defined as
\begin{figure}[t!]
	\vspace*{0.3mm}
	\centering
	\includegraphics[width=0.49\textwidth]{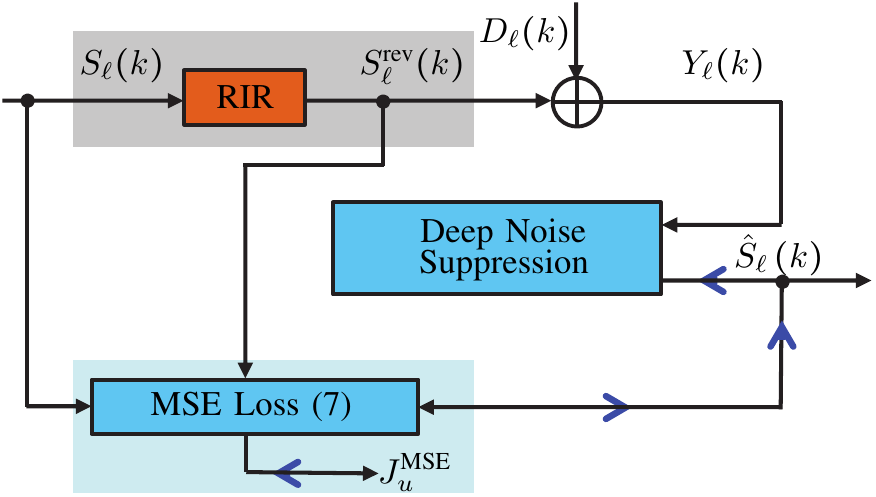}
	\caption{{\bf DNS pre-training setup.} The blue arrows indicate the gradient flow back-propagated for the DNS model pre-training.}
	\label{system2}
\end{figure}
%%%%%%%%%%%%%%%%%%%%%%%%%%%%%%%%%%%%%%%%%
\begin{figure}[t!]
	\centering
	\includegraphics[width=0.49\textwidth]{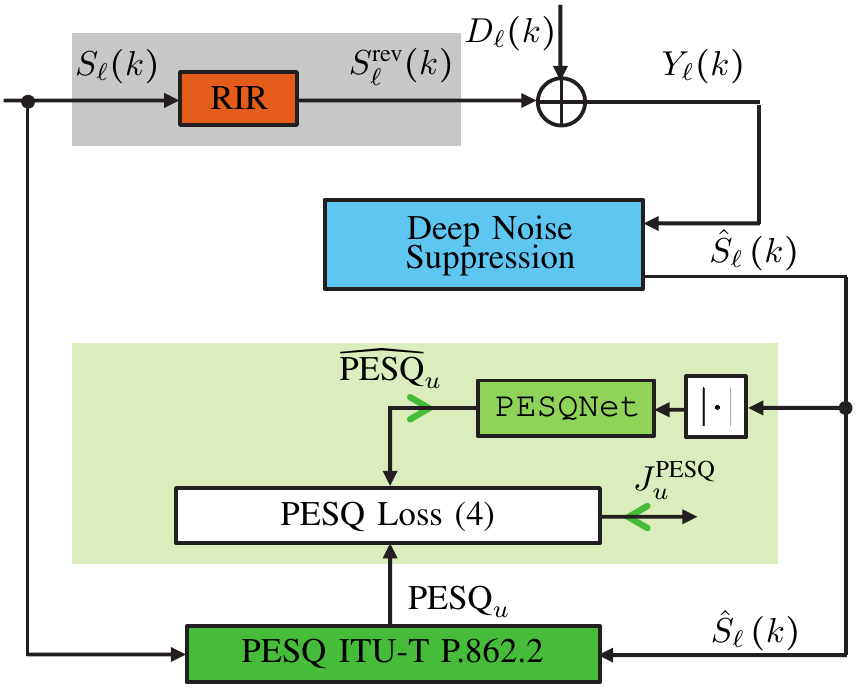}
	\caption{{\bf \texttt{PESQNet} pre-training setup.} The deep noise suppression is fixed, and the green arrows indicate the gradient flow back-propagated for the {\tt PESQNet} model pre-training.}
	\label{system6}
\end{figure}
%%%%%%%%%%%%%%%%%%%%%%%%%%%%%%%%%%%%%%%%%%%
\begin{equation} \label{joint}
J^\text{joint}_u\!=\!\frac{1}{L_u\!\cdot\!K}\!\sum_{\ell\in\mathcal{L}_u}\sum_{k\in\mathcal{K}}\!\bigl|\hat{S}_\ell(k)\!-\!S_\ell(k)\bigr|^2,
\end{equation}
with $\mathcal{L}_u$ being the set of frame indices of the utterance indexed with $u$, and $L_u\!=\!\bigl|\!\mathcal{L}_u\!\bigr|$ being its number of frames. The second utterance-wise loss term only focuses on denoising by employing the reverberated clean speech spectrum $S^\text{rev}_\ell(k)$ as target, following
\begin{equation} \label{noise}
J^\text{noise}_u\!=\!\frac{1}{L_u\!\cdot\!K}\!\sum_{\ell\in\mathcal{L}_u}\sum_{k\in\mathcal{K}}\!\bigl|\hat{S}_\ell(k)\!-\!S^\text{rev}_\ell(k)\bigr|^2\!.
\end{equation}
Afterwards, the two loss terms \eqref{joint} and \eqref{noise} are combined in the joint loss function as:
\begin{equation} \label{MT}
J^\text{MSE}_u\!=\beta\cdot J^\text{joint}_u+(1-\beta)\cdot J^\text{noise}_u,
\end{equation}
with $\beta\in\left [ 0,1 \right ]$ being the weighting factor to control a weaker or a stronger dereverberation, where $\beta=0$ lets \eqref{MT} become the conventional MSE loss for pure denoising. The pre-training setup for DNS employing \eqref{MT} is illustrated in Fig.\,\ref{system2}, where the upper gray box illustrates the microphone signal model, including the room impulse response (RIR), while the lower part contains the computation of loss \eqref{MT}. The blue arrows indicate the gradient flow from the MSE loss \eqref{MT} back-propagated to the DNS for pre-training. The details of the employed DNS model are illustrated in Fig.\,\ref{fig:CNN_topology}.
\begin{figure}[t!]
	\vspace*{0.5 mm}
	\centering
	\includegraphics[width=0.49\textwidth]{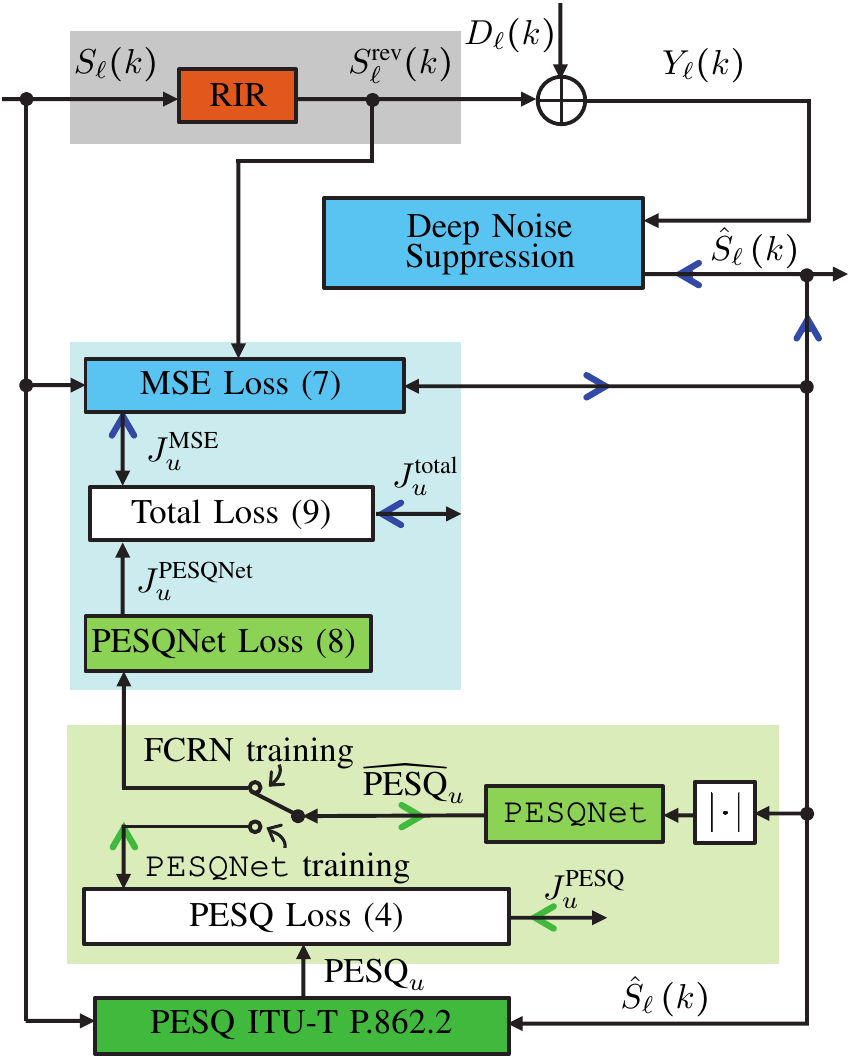}
	\caption{{\bf \texttt{PESQNet} and DNS training setup.} After the pre-trainings, the DNS and the {\tt PESQNet} are trained alternately, controlled by the switch in upper and lower position, respectively. The total loss $J^\text{total}_u$ \eqref{total} in the center controls the DNS training mediated by a fixed {\tt PESQNet}. The lower part shows the {\tt PESQNet} training adapting it to the current fixed DNS, employing $J^\text{PESQ}_u$ \eqref{PESQNet}. The blue and green arrows indicate the gradient flow back-propagated for DNS and {\tt PESQNet} training, respectively.}
	\label{system3}
\end{figure}

\subsubsection{PESQNet Pre-Training}
Afterwards, the DNS is fixed and the {\tt PESQNet} is trained to adapt to the current pre-trained DNS employing \eqref{PESQNet}. This is illustrated in the {\tt PESQNet} pre-training setup in Fig.\,\ref{system6}, where the green arrows indicate the gradient flow back-propagated for the {\tt PESQNet} pre-training. The details of the employed {\tt PESQNet} are illustrated in Fig.\,\ref{fig:PESQNet}.

\subsubsection{Fine-Tuning}
In the fine-tuning stage as it is shown in Fig.\,\ref{system3}, the pre-trained {\tt PESQNet} is applied to the output of the pre-trained DNS, estimating the PESQ scores of the enhanced speech. Thus, it serves as a differentiable PESQ loss for the training of the DNS, aiming at maximizing the PESQ of the output enhanced speech signal. Accordingly, we can define a ``PESQNet loss"  
\begin{equation} \label{real}
J^\text{PESQNet}_u\!=\left(\widehat{\text{PESQ}}_u-\text{PESQ}_\text{max}\right)^2
\end{equation}
for utterance $u$, with $\text{PESQ}_\text{max}\!=\!4.64$, which is minimized during DNS training. Furthermore, we explicitly consider joint dereverberation and denoising also in the fine-tuning by combining the joint MSE-based loss \eqref{MT} with the novel reference-free PESQNet loss \eqref{real}. Thus, the total loss function employed during fine-tuning of the DNS is defined as
\begin{equation} \label{total}
J^\text{total}_u= \alpha\cdot J^\text{MSE}_u\!+\!(1\!-\!\alpha)\!\cdot\!J^\text{PESQNet}_u,
\end{equation}
with $\alpha\in\left [ 0,1 \right ]$ being the weighting factor. By choosing $\alpha$ close to $0$, the novel reference-free psychoacoustic loss $J^\text{PESQNet}_u$ will dominate the DNS fine-tuning, and is supposed to deliver a better perceptual speech quality.

One major contribution of this work is to solve the issues addressed in \cite{fu2019learning}, where the fixed {\tt PESQNet} was reported to be fooled by the updated DNS (estimated PESQ scores increase while true PESQ scores decrease) after training for several minibatches. In \cite{fu2019learning}, this was mainly caused by the fixed {\tt Quality-Net} not having seen the enhanced speech signal generated by the updated DNS. Accordingly, we propose to train the DNS and the {\tt PESQNet} alternatingly on an {\it epoch} level to keep the {\tt PESQNet} up-to-date, and most importantly: it specifically adapts to the DNS in its current learning step. In this novel alternating training protocol, the DNS and the employed {\tt PESQNet} are trained with the loss \eqref{total} and \eqref{PESQNet}, respectively, as illustrated in Fig.\,\ref{system3}.

In Fig.\,\ref{system3}, the alternating training for the DNS and the {\tt PESQNet} is controlled by the switch in upper and lower position, respectively. The middle part containing the total loss $J^\text{total}_u$ \eqref{total} computation denotes the DNS training controlled by a fixed {\tt PESQNet}. The lower part shows the {\tt PESQNet} training adapting it to the current fixed DNS, employing $J^\text{PESQ}_u$ \eqref{PESQNet}. The blue and green arrows indicate the gradient flow back-propagated for DNS and {\tt PESQNet} training, respectively. The detailed structures of the DNS as well as the employed {\tt PESQNet} are illustrated in Figs.\,\ref{fig:CNN_topology} and \ref{fig:PESQNet}, respectively.
%%%%%%%%%%%%%%%%%%%%%%%%%%%%%%%%%%%%%%%%%%%%%%%%%%%%%%%%%%%%%%%%%%%%%%%%%%%%%%%%%%%%%%%%%
\section{Experimental Setup and Databases}
\subsection{Dababase and Preprocessing}
We perform a two-step training, which includes pre-training and fine-tuning steps. During pre-training, the dataset $\mathcal{D}_\text{WSJ0}$ comprises a $105$-hours training set $\mathcal{D}^\text{train}_\text{WSJ0}$ and an $18$-hours validation set $\mathcal{D}^\text{val}_\text{WSJ0}$, which are synthesized with the clean speech from WSJ0 speech corpus \cite{Garofalo2007} and noise from DEMAND \cite{thiemann2013diverse} and QUT \cite{dean2010qut} comprising 35 different noise files shared in training and validation. We normalized the clean speech active speech level to $-26\,\text{dBov}$ and simulated five SNR conditions ranging from $0$ to $20$ dB with a step size of $5$ dB, according to ITU-T P.56 \cite{ITU56}. Please note that no reverberation effects are considered in preparation of the pre-training dataset. Furthermore, a small test set $\mathcal{D}^\text{test}_\text{WSJ0}$ is prepared by mixing the clean speech from eight unseen speakers with four unseen types of noise taken from DEMAND \cite{thiemann2013diverse} and QUT \cite{dean2010qut}, including SNRs ranging from $0$ to $10$ dB with a $5$ dB step size. The $\mathcal{D}^\text{test}_\text{WSJ0}$ dataset is only used to evaluate the {\tt PESQNet} performance after pre-training.

The fine-tuning dataset contains files randomly chosen from the official Interspeech 2021 DNS Challenge (dubbed DNS3) training material \cite{reddy2021interspeechIN}. This $\mathcal{D}_\text{DNS3}$ dataset contains $100$ hours of training material $\mathcal{D}^\text{train}_\text{DNS3}$ and $10$ hours of validation material $\mathcal{D}^\text{val}_\text{DNS3}$, where SNRs are sampled uniformly between $0$ and $40$ dB. The RMS level of the mixture is set to a value uniformly sampled between $-38$ and $-18$ dBov. The organizers also provided both recorded and synthetic RIRs in the dataset \cite{reddy2021icassp}. However, we do not use their provided RIRs, since the time shift caused by the provided RIRs is unknown, so that the training input and the corresponding target are not time-aligned, which is problematic for training a complex mask-based DNS. Thus, we reverberated $50\%$ of the files in $\mathcal{D}_\text{DNS3}$ by convolving the clean speech component with our simulated RIRs. The employed RIRs are simulated using the mirror method \cite{scheibler2018pyroomacoustics} with the room size uniformly sampled from  ({\it length, width, height}) $\in$ ([$3, 10$] m, [$3, 10$] m, [$2.5, 3.5$] m) and an absorption coefficient uniformly sampled between $0.1$ and $0.3$ for all room surfaces. The microphone is assumed in the center of the simulated room and the source is randomly placed at a source-to-microphone distance uniformly sampled between $0.1$ m and $1$ m. Accordingly, the estimated RT60s lie between $0.28$ and $1.66$\,s.

We use the preliminary synthetic test set from the first Interspeech 2020 DNS Challenge (DNS1) \cite{reddy2020interspeech} for development (dubbed $\mathcal{D}^\text{dev}_\text{DNS1}$), which contains $150$ noisy speech mixtures with and without reverberation. Please note that we use this dataset $\mathcal{D}^\text{dev}_\text{DNS1}$ for instrumental performance measurement, whereas the preliminary synthetic test set from the DNS3 challenge \cite{reddy2021interspeechIN} contains target signals of singing and emotions (e.g., crying, yelling, and laughing). Accordingly, it would be inadequate to evaluate the speech enhancement performance employing instrumental metrics such as PESQ \cite{ITUT_pesq_wb_corri} and STOI \cite{taal2010short}, which are designed for speech signals only.

For the final evaluation, we prepare both {\it synthetic} and {\it real} test datasets. The {\it synthetic} test set contains all speech files of the official preliminary test set of the ICASSP 2020 DNS Challenge (dubbed DNS2) \cite{reddy2021icassp}. This {\it synthetic} test dataset denoted as $\mathcal{D}^\mathrm{test}_\mathrm{DNS2}$ is used for instrumental performance measurements.  To further evaluate our proposed methods in real implementations, we use the preliminary test set from DNS3 \cite{reddy2021interspeechIN} denoted as $\mathcal{D}^\mathrm{test}_\mathrm{DNS3}$ as our {\it real} test set, which contains noisy speech recorded in reality. 

In this work, signals have a sampling rate of $16\, \text{kHz}$ and we apply a periodic Hann window with frame length of $384$ with a $50\%$ overlap, followed by an FFT with $K=512$. The number of input and output frequency bins in Figs.\,\ref{fig:CNN_topology} and \ref{fig:PESQNet} is set to $K_{\rm in}=260$. The last 3 frequency bins are redundant and only used for compatibility with the two maxpooling and upsampling operations in the employed DNS model shown in Fig.\,\ref{fig:CNN_topology}, and are dropped for subsequent processing. Since we perform a complex mask-based speech enhancement, the number of input and output channels in Fig.\,\ref{fig:CNN_topology} represented by $C_{\rm in}$ and $C_{\rm out}$ are set to $2$, reflecting the real and the imaginary parts.
\subsection{Training Protocols}
\subsubsection{DNS Pre-Training}
Firstly, we pre-train the DNS model using $\mathcal{D}^\text{train}_\text{WSJ0}$ with the loss \eqref{MT}. The DNS pre-training setup is illustrated in Fig.\,\ref{system2}, where the blue arrows indicate the gradient flow back-propagated for the DNS model pre-training. Since no reverberation effects are considered in the pre-training stage, $S_\ell(k)=S^\text{rev}_\ell(k)$ holds in Fig.\,\ref{system2}. Accordingly, we set $\beta=0$ in \eqref{MT}. For our employed {\tt FCRN} shown in Fig.\,\ref{fig:CNN_topology}, the number of filter kernels is set to $F=88$ and the kernel size is chosen as $N=24$. In both pre-training and fine-tuning for our DNS, we employ a truncated backpropagation-through-time (BPTT) training with a sequence length (unrolling depth of BPTT) equal to the number of time frames belonging to the current input utterance. Furthermore, a batch size of 3 utterances is employed. For the DNS pre-training, we employ the Adam optimizer with the initial learning rate of $10^{-4}$. The learning rate is halved once the validation loss measured on $\mathcal{D}^\text{val}_\text{WSJ0}$ does not improve for a consecutive five epochs. We stop the training after the learning rate is decreased below $10^{-5}$, and the DNS model which provides the lowest validation loss is saved. 
\subsubsection{PESQNet Pre-Training}
Secondly, the {\tt PESQNet} is pre-trained on the same dataset with the enhanced speech spectrum $\hat{S}_\ell(k)$ generated by the pre-trained and fixed {\tt FCRN} as illustrated in Fig.\,\ref{system6}. The training targets are the corresponding ground truth PESQ scores of the enhanced speech signals measured with the original ITU-T P.862.2 PESQ function \cite{ITUT_pesq_wb_corri}, as illustrated in the lower part in Fig.\,\ref{system6}. As in DNS pre-training, we have $S_\ell(k)=S^\text{rev}_\ell(k)$ in Fig.\,\ref{system6}. For our employed {\tt PESQNet} shown in Fig.\,\ref{fig:CNN_topology}, the widths of the convolutional kernels are set to $w_i=2^{i-1}$, $i\in\left\{1,2,3,4\right\}$. The number of time frames for the input feature matrices shown in Fig.\,\ref{fig:PESQNet} is set to $W=16$. In both pre-training and fine-tuning of our {\tt PESQNet}, we employ the same truncated BPTT training scheme as used for DNS training. The Adam optimizer is employed with an initial learning rate of $2\!\cdot\!10^{-4}$. The learning rate is halved once the validation loss does not improve for five consecutive epochs. We stop the training when the learning rate is smaller than $10^{-5}$ and the model with the lowest validation loss is saved.
\subsubsection{Novel FCRN and PESQNet Alternating Fine-Tuning}
We propose a novel two-stage fine-tuning protocol on the dataset $\mathcal{D}_\text{DNS3}$ from DNS3. In the first stage, which is basically a domain adaptation, we fine-tune the pre-trained {\tt FCRN} with the loss \eqref{MT} on the $\mathcal{D}^\text{train}_\text{DNS3}$ dataset and subsequently, still in the first stage, we fine-tune the {\tt PESQNet} on the same dataset with the enhanced speech spectrum generated by the fixed fine-tuned {\tt FCRN}, again with loss \eqref{PESQNet}. During fine-tuning, $50\%$ of the files in $\mathcal{D}_\text{DNS3}$ contain reverberation, therefore, we set $\beta=0.9$ in \eqref{MT} for joint denoising and dereverberation as proposed by \cite{strake2020DNS}. The initial learning rates for the training of the DNS and the {\tt PESQNet} are set to $2\!\cdot\!10^{-5}$ and $5\!\cdot\!10^{-5}$, respectively. Both of the two trainings are stopped when the respective learning rate is smaller than $10^{-6}$. Other settings are exactly the same as employed in its corresponding pre-training.

In the second-stage (final) fine-tuning, we alternatingly fine-tune the DNS with the fixed {\tt PESQNet} serving as mediator towards ITU-T P.862.2, and fine-tune the {\tt PESQNet} with a fixed DNS.

A {\it cycle} of our novel alternating training protocol is defined as follows: We first train the {\tt FCRN} with a fixed {\tt PESQNet} using the total loss \eqref{total} on {\it one epoch} of training data from $\mathcal{D}^\text{train}_\text{DNS3}$. This is illustrated in Fig.\,\ref{system3} with the switch in the upper position. Then, we fix the {\tt FCRN}, and train the {\tt PESQNet} also on {\it one epoch} of training data from the same dataset using the PESQ loss \eqref{PESQNet}, with the enhanced speech obtained from the current fixed {\tt FCRN}. This is shown in Fig.\,\ref{system3} with the switch changed to the lower position. This training protocol is dubbed as $\langle1\!-\!1\rangle$, where ``1" denotes training with one epoch of data from $\mathcal{D}^\text{train}_\text{DNS3}$. In this second-stage fine-tuning, the DNS and the {\tt PESQNet} are trained with in total $25$ epochs of data, with epoch index $\tau\in\left\{1,2,\ldots,25\right\}$. Here, the DNS is trained with a fixed {\tt PESQNet} on epochs with odd index number, e.g., $\tau\in\left\{1,3,5,\ldots,25\right\}$, while the {\tt PESQNet} is trained adapting to the fixed updated DNS on epochs with even index number. We fix the learning rates for the training of the DNS and the {\tt PESQNet} to $10^{-6}$ and $2\!\cdot\!10^{-6}$, respectively. The DNS model with the lowest validation loss as well as its corresponding {\tt PESQNet} are saved.

To explore the influence of the hyperparameter $\alpha$ in \eqref{total} on the DNS second-stage fine-tuning, we employ $\alpha\in\left \{0, 0.5, 1\right \}$. Please note that for the hyperparameter setting $\alpha=1$, the DNS training is not controlled by our proposed {\tt PESQNet}. In this case, the loss functions used in the first-stage and the second-stage fine-tuning are exactly the same. Thus, this training serves as a ``placebo" setup intended to prove that the perceptual quality improvements for $\alpha\!<\!1$ are not just caused by the extra second-stage fine-tuning with much lower learning rate.  

Our proposed alternating training protocol resembles the alternating training schemes used in adversarial trainings, which are known to have issues in training stability \cite{shrivastava2017learning,miyato2018spectral}. Such instable training can usually be observed by a fluctuating or increasing loss measured even on the training dataset $\mathcal{D}^\text{train}_\text{DNS3}$. To further stabilize the second-stage fine-tuning, we employ a simple accumulating of gradients during the DNS fine-tuning, where the gradients obtained from every minibatch of training data within one epoch are accumulated and averaged for the final weight update at the end of the epoch. Please note that this gradient accumulation is only applied to the DNS second-stage fine-tuning, not for the {\tt PESQNet} second-stage fine-tuning. Compared to the DNS pre-training and the first-stage fine-tuning, the weights of the DNS model employing gradient accumulation are only updated once per epoch.
\subsection{Baselines}
\subsubsection{DNS Trained with MSE-Based Loss}
As one of the baselines, we employ the MSE-based loss \eqref{MT} proposed in \cite{strake2020DNS} to train our DNS. Strake et al.\ \cite{strake2020DNS} trained the same DNS model shown in Fig.\,\ref{fig:CNN_topology} with this MSE-based denoising and dereverberation loss in the DNS1 Challenge \cite{reddy2020interspeech} and achieved a second rank in the non-realtime track, although their model (as ours) in their work are actually realtime-capable.

Indeed, this strong baseline DNS is exactly the one obtained from our first-stage fine-tuning, since up to this stage even datasets match exactly. Accordingly, by comparing to this baseline, we intend to exploit how much perceptual speech quality could be gained by employing the additional second-stage fine-tuning mediated by our proposed {\tt PESQNet}. This baseline is denoted as ``{\tt FCRN} \cite{strake2020DNS}" in the following discussions.
\subsubsection{Microsoft DNS3 Challenge Baseline}
As another reference, we take the baseline provided by Microsoft \cite{braun2020data} in the DNS3 Challenge \cite{reddy2021interspeechIN}. Braun et al.\ proposed a level-invariant normalized loss function to train a recurrent neural network (RNN) for a complex mask-based speech enhancement. The proposed loss function is supposed to avoid signals with high active levels dominating the training process. Furthermore, data augmentation techniques are employed by considering different SNR levels, speech active levels, and also filtering effects caused by acoustics or recording devices. 

Please note that we directly employ the realtime-capable, fully-trained DNS3 baseline as provided by the challenge organizers \cite{reddy2021interspeechIN} without any re-training since we anyway evaluate on DNS Challenge data. Thus, a fair comparison can be made between our work and the DNS3 baseline \cite{braun2020data}, based on the DNS3 challenge rules specified in \cite{reddy2021interspeechIN}. Furthermore, we call it ``DNS3 Baseline \cite{reddy2021interspeechIN}" in this work. 
\subsubsection{Baseline from Our Previous Work}
In our previous work \cite{xu2021inter}, we had proposed to use this end-to-end non-intrusive {\tt PESQNet} to advantageously employ {\it real recordings} in DNS training. This is achieved by a ``weakly" supervised training with both synthetic and real data on {\it minibatch level}. In \cite{xu2021inter}, the novelty was to focus on employing real training data in training a speech enhancement DNN without using GAN-type losses. However, the training of \cite{xu2021inter} without employing gradient accumulation was sensitive w.r.t.\ instabilities, thus, very limited performance improvement was achieved for real test data, and even no PESQ improvement was achieved for synthetic test data. Solving the instability problem reported in \cite{fu2019learning} and sometimes occurring in \cite{xu2021inter}, is a core contribution of this work, along with a much more extensive analysis and experimental evaluation. Therefore, we adopt our previous work as another baseline to show the benefits obtained from the novel $\langle1\!-\!1\rangle$ epoch-wise training protocol employing gradient accumulation. Our previous work is represented as ``{\tt FCRN}/{\tt PESQNet} \cite{xu2021inter}" in the following result tables.
\subsection{Quality Metrics}
To evaluate the performance of the DNS model, we employ measurements with instrumental metrics such as PESQ \cite{ITUT_pesq_wb_corri}, STOI \cite{taal2010short}, segmental SNR improvement $\Delta\text{SNR}_\text{seg}$ \cite{loizou2013speech}, and speech-to-reverberation modulation energy ratio (SRMR) \cite{falk2010non}. PESQ and STOI are measured on the enhanced speech signal reflecting the perceptual speech quality and intelligibility, respectively. For the noisy mixtures without reverberation, we measure the $\Delta\text{SNR}_\text{seg}$ according to \cite{loizou2013speech} to explicitly evaluate the denosing effects. SRMR is measured only on the noisy mixtures under reverberated conditions to evaluate the dereverberation effects. Furthermore, we also measured DNSMOS \cite{reddy2021dnsmos} on the enhanced speech, which is obtained from a non-intrusive network specifically trained to predict human subjective rating scores for DNS tasks according to ITU-T P.808 \cite{ITU_P808}. All of the abovementioned metrics should be as high as possible.

The performance of the {\tt PESQNet} is reported by the mean absolute error (MAE) and the linear correlation coefficient (LCC) as used in \cite{fu2018quality}. Both of the metrics are calculated based on the estimated PESQ score from the {\tt PESQNet} and its corresponding ground truth measured according to ITU-T P.862.2 PESQ \cite{ITUT_pesq_wb_corri}. An accurately estimated PESQ score is reflected by a low MAE and a high LCC.
%%%%%%%%%%%%%%%%%%%%%%%%%%%%%%%%%%%%%%%%%%%%%%%%%%%%%%%%%%%%%%%%%%%%%%%%% 
\section{Experimental Results and Discussion}
\subsection{PESQNet Performance}
Since the proposed {\tt PESQNet} will serve to mediate towards the \nth{2}-stage fine-tuning of the DNS, an accurately estimated $\widehat{\text{PESQ}}_u$ employed in the total loss \eqref{total} is crucial in guiding the DNS to further increase the perceptual quality. Therefore, we have to put focus on the performance of the {\tt PESQNet} after the pre-training and after the \nth{1}-stage fine-tuning.
\begin{figure}[t!]
	\centering
	\includegraphics[width=0.48\textwidth]{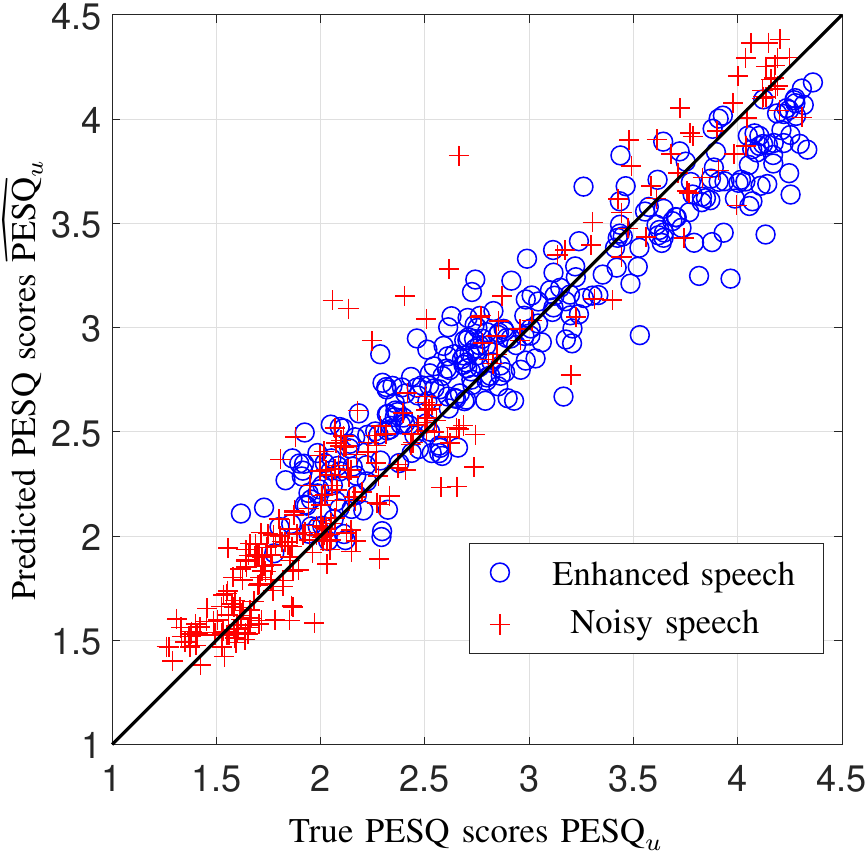}
	\caption{Scatter plot for the predicted PESQ scores $\widehat{\text{PESQ}}_u$ by {\tt PESQNet}, measured on $\mathcal{D}^\text{test}_\text{WSJ0}$, {\bf after pre-training} of first the {\tt FCRN} and then the {\tt PESQNet} on $\mathcal{D}^\text{train}_\text{WSJ0}$. The enhanced speech signal $\hat{s}(n)$ used for predicting $\text{PESQ}_u$ is obtained from the corresponding {\bf pre-trained \texttt{FCRN}}.}
	\label{PESQNet_pre}
\end{figure}
%%%%%%%%%%%%%%%%%%%%%%%%%%%%%%%%%%%%%%%%%%%%%%%%%%%%%%%%%%%%%%%%%%
\begin{table}[t]
	%\large
	\centering
	\caption{Performance of the {\tt PESQNet} measured on $\mathcal{D}^\text{test}_\text{WSJ0}$, {\bf after} {\bf pre-training} on $\mathcal{D}^\text{train}_\text{WSJ0}$. Performance is measured using the mean absolute error (MAE) and the linear correlation coefficient (LCC), between the ground truth PESQ and predicted PESQ scores. The enhanced speech signal $\hat{s}(n)$ used for predicting $\text{PESQ}_u$ is obtained from the corresponding {\bf pre-trained \texttt{FCRN}}.}
	\setlength\tabcolsep{4pt}
	%\resizebox{1\linewidth}{!}{
	\begin{tabular}{c c c c c}
		\hline
		&\stz{MAE($\hat{s}$)} & MAE($y$) & LCC($\hat{s}$)& LCC($y$)\\ \cmidrule(r){2-3} \cmidrule(r){4-5}
		\rotatebox{90}{}	& \stz{0.21} & \stz{0.16} & \stz{0.95}& \stz{0.97}\\ \hline
		
	\end{tabular}
	%}
	\label{PESQ_pretrain}
\end{table}
%%%%%%%%%%%%%%%%%%%%%%%%%%%%%%%%%%%%%%%%%%%%%%%%%%
\begin{figure}[t!]
	\centering
	\includegraphics[width=0.48\textwidth]{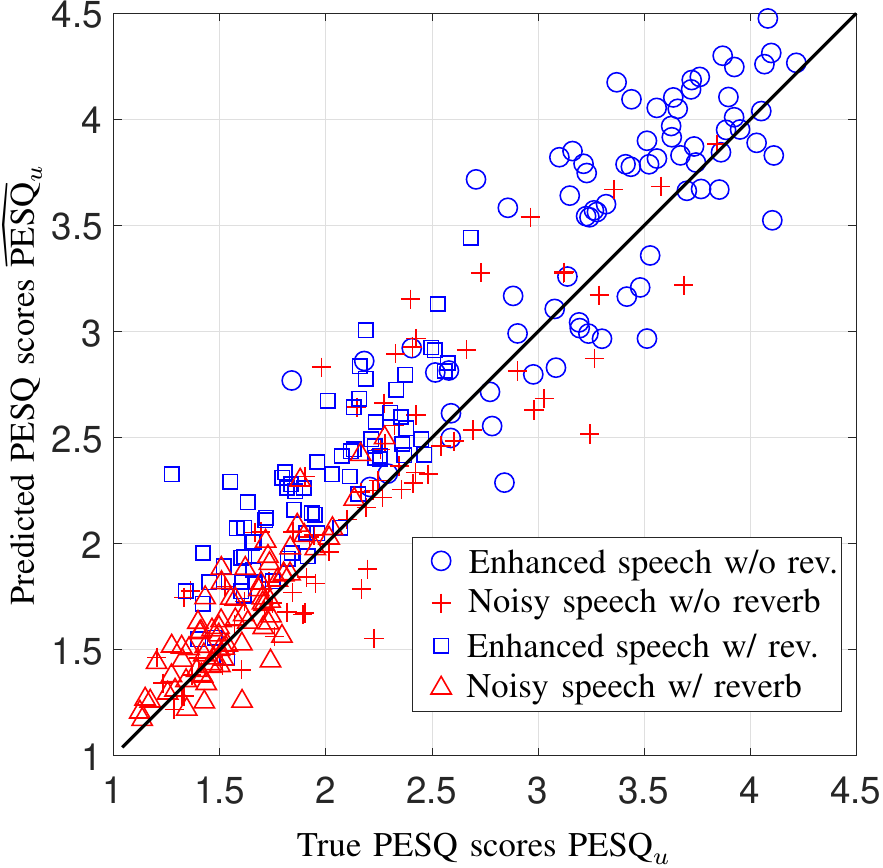}
	\caption{Scatter plot for the predicted PESQ scores $\widehat{\text{PESQ}}_u$ by {\tt PESQNet}, measured on $\mathcal{D}^\mathrm{dev}_\mathrm{DNS1}$, {\bf after first-stage fine-tuning} of first the {\tt FCRN} and then the {\tt PESQNet} on $\mathcal{D}^\text{train}_\text{DNS3}$ \cite{reddy2021interspeechIN}. The enhanced speech signal $\hat{s}(n)$ used for predicting $\text{PESQ}_u$ is obtained from the corresponding {\bf first-stage fine-tuned \texttt{FCRN}}. The performance is reported on both reverberation conditions.}
	\label{PESQNet_fine}
\end{figure}
%%%%%%%%%%%%%%%%%%%%%%%%%%%%%%%%%%%%%%
\begin{table}[t!]
	%\large
	\centering
	\caption{Performance of the {\tt PESQNet} measured on $\mathcal{D}^\mathrm{dev}_\mathrm{DNS1}$, {\bf after first-stage fine-tuning} on the DNS3 training set $\mathcal{D}^\text{train}_\text{DNS3}$ \cite{reddy2021interspeechIN}. Performance is measured using the mean absolute error (MAE) and the linear correlation coefficient (LCC), between the ground truth PESQ and predicted PESQ scores. The enhanced speech signal $\hat{s}(n)$ used for predicting PESQ($\hat{s}$) is obtained from the corresponding {\bf first-stage fine-tuned \texttt{FCRN}}. The performance is reported on both reverberation conditions.}
	\setlength\tabcolsep{2pt}
	%\resizebox{1\linewidth}{!}{
	\begin{tabular}{c c c c c}
		\hline
		&\stz{MAE($\hat{s}$)} & MAE($y$) & LCC($\hat{s}$)& LCC($y$)\\ \cmidrule(r){1-3} \cmidrule(r){4-5}
		Reverberation	& \stz{0.32} & \stz{0.12} & \stz{0.85}& \stz{0.87}\\ \hline
		No reverberation	& \stz{0.31} & \stz{0.23} & \stz{0.82}& \stz{0.88}\\ \hline
		
	\end{tabular}
	%}
	\label{PESQ_finetune}
\end{table}
\subsubsection{PESQNet Pre-Training Performance}
After the {\tt PESQNet}'s pre-training illustrated in Fig.\,\ref{system6} with $\mathcal{D}^\text{train}_\text{WSJ0}$, we measure its performance on $\mathcal{D}^\text{test}_\text{WSJ0}$. Please note that we do not consider reverberations in the pre-training and the following performance evaluation. 

In Fig.\,\ref{PESQNet_pre}, we present the scatter plot with the predicted PESQ score $\widehat{\text{PESQ}}_u$ obtained from the pre-trained {\tt PESQNet} and its ground truth $\text{PESQ}_u$ measured by ITU-T P.862.2 PESQ \cite{ITUT_pesq_wb_corri}. Noisy speech utterances and their enhanced version from the corresponding pre-trained DNS are represented by red and blue markers, respectively. We can see that the PESQ scores of the enhanced speech utterances are slightly more challenging to predict than the noisy ones, which is reflected by more blue markers distributed away from the diagonal line. The measured MAE and LCC in Table\,\ref{PESQ_pretrain} confirm our observation, where we achieve $0.16$ MAE and $0.97$ LCC for the noisy speech, which is slightly better than the values measured on enhanced speech. However, we can generally see that the pre-trained {\tt PESQNet} is quite reliable, especially from the high LCC (around $0.96$) and the low MAE (around 0.19) averaged over both noisy and enhanced speech, shown in Table\,\ref{PESQ_pretrain}.
\subsubsection{PESQNet \nth{1}-Stage Fine-Tuning Performance}
After the \nth{1}-stage fine-tuning of {\tt PESQNet} with $\mathcal{D}^\text{train}_\text{DNS3}$, we evaluate its performance on $\mathcal{D}^\mathrm{dev}_\mathrm{DNS1}$. Please note that we consider the conditions with and without reverberation in the fine-tuning and the subsequent performance evaluation. 

In Fig.\,\ref{PESQNet_fine}, we present the scatter plot with the same metrics used in Fig.\,\ref{PESQNet_pre} for the noisy and the enhanced speech utterances, however, under both reverberation conditions.  We can see that the distribution of the markers is more dispersive and farther away from the diagonal when compared to the ones in Fig.\,\ref{PESQNet_pre}. This is supposedly caused by the increased diversities in the training and the test datasets, w.r.t.\ noise types, languages, reverberations, SNRs, and speech active levels, leading to a more challenging task. In Table\,\ref{PESQ_finetune}, the MAE values measured on the enhanced speech are around $0.1$ points higher than that in Table\,\ref{PESQ_pretrain}. Furthermore, we can see that an overall performance degradation on both noisy and enhanced speech reflected by around $0.1$ points lower LCC values compared to the ones in Table\,\ref{PESQ_pretrain}. However, in general, the fine-tuned {\tt PESQNet} is still reliable by offering an LCC above $0.85$, averaged over the noisy and the enhanced speech under both reverberation conditions. Furthermore, the performance of the {\tt PESQNet} is very balanced over both reverberation conditions reflected by the similar LCC values, which is important for mediating towards the \nth{2}-stage fine-tuning of the DNS.
%%%%%%%%%%%%%%%%%%%%%%%%%%%%%%%%%%%%%%%%%%%%%%%%%%%%%%%%%%%%%%%%
\subsection{DNS \nth{2}-Stage Alternating Fine-Tuning Results}
%%%%%%%%%%%%%%%%%%%%%%%%%%%%%%%%%%%%%%%%%%%%%%%%%%%%%%%%%%%%%%%%%%%%%%%%
\begin{table*}[!t]
	\centering
	\caption{\textbf{Instrumental quality results} on the \textbf{development set} $\mathcal{D}^\mathrm{dev}_\mathrm{DNS1}$, employing {\bf synthetic data}. Evaluation is performed separately on the conditions without and with reverberation and is used for the optimization of hyperparameter $\alpha$ in \eqref{total}. DNSMOS is adopted from \cite{reddy2021dnsmos}. Best results are in {\bf bold} font, and the second best are \underline{underlined}. Proposed method with $*$.}
	\setlength\tabcolsep{4pt}
	%\resizebox{0.98\linewidth}{!}{
		\begin{tabular}{ccc c c c c c c c}
			\hline
			&\multirow{2}{*}{Methods} & \multicolumn{4}{c}{\stz{Without reverb}}& \multicolumn{4}{c}{\stz{With reverb}}\\ \cmidrule(r){3-6} \cmidrule(r){7-10} %\hhline{~~--------}
			&    &  PESQ & DNSMOS  &  \stz{STOI}  &   $\Delta\text{SNR}_\text{seg}$[dB]   & PESQ & DNSMOS   & \stz{STOI} &  SRMR \\ \hhline{----------} 
			\multirow{4}{*}{\rotatebox{90}{REF\ }}& \multicolumn{1}{l}{Noisy}             &  \stz{2.21} &  3.15  &  0.91   &  -  &  1.57  &  2.73  &0.56    &-  \\ \hhline{~~~~~~~~~~}
			& \multicolumn{1}{l}{\stz{DNS3 Baseline \cite{braun2020data}}}              &     3.15   &  3.64    &    0.94   &    {6.30} & 1.68   &  {\bf 3.18}   & \underline{0.62}   & 6.33\\ \hhline{~~~~~~~~~~}
			& \multicolumn{1}{l}{\stz{{\tt FCRN} \cite{strake2020DNS}}}  &      {3.37}   &  {3.82}   &    {\bf 0.96}   &    8.35 & {\bf 1.95} &  3.08    &  {\bf 0.63}  & {7.25}\\ \hhline{~~~~~~~~~~}
			&\multicolumn{1}{l}{\stz{{\tt FCRN}/{\tt PESQNet}} \cite{xu2021inter}}   &    {3.35}  &  {\bf 3.88}    &   \underline{0.95}  &    {8.40}  & {1.92} &  3.11   &  {0.61}  & {\bf 7.41}  \\ \hhline{----------}
			\multirow{3}{*}{\rotatebox{90}{NEW}} &\multicolumn{1}{l}{\stz{{\tt FCRN}/{\tt PESQNet}$^*$, $\langle1\!-\!1\rangle$, $\alpha\!=\!0$}}   &    {\bf 3.45}  &  \underline{3.87}  & {\bf 0.96}  &    {\bf 8.48}  & {\bf 1.95}  &  \underline{3.13}   &  \underline{0.62}  & 7.38 \\ \hhline{~~~~~~~~~~}
			&\multicolumn{1}{l}{\stz{{\tt FCRN}/{\tt PESQNet}, $\langle1\!-\!1\rangle$, $\alpha\!=\!0.5$}}   &    \underline{3.43}  &  {3.86}   &    {\bf 0.96}  & \underline{8.42}  & {\bf 1.95}  &  3.12   &  \underline{0.62}  & \underline{7.40} \\ \hhline{~~~~~~~~~~}
			&\multicolumn{1}{l}{\stz{{\tt FCRN}/{\tt PESQNet}, $\langle1\!-\!1\rangle$, $\alpha\!=\!1$}}   &    3.36  &  3.82   &   {\bf 0.96}  & 8.34  &     \underline{1.94}  &  3.09   &  {\bf 0.63}  & 7.25 \\ \hhline{----------}
		\end{tabular}
	%}
	\label{DNS1_dev}
\end{table*}
\subsubsection{Hyperparameter Optimization and Analysis}
To explore the influence of the hyperparameter $\alpha$ in \eqref{total}, we employ $\alpha\in\left \{0, 0.5, 1\right \}$ in the DNS \nth{2}-stage fine-tuning. Afterwards, we evaluate the performance of our trained DNS models and the baseline methods on the synthetic dataset $\mathcal{D}^\mathrm{dev}_\mathrm{DNS1}$. The performance is reported separately on the conditions with and without reverberation as shown in Table\,\ref{DNS1_dev}. The best results are marked in {\bf bold} font and the second best are \underline{underlined}.

It can be seen that among all the baseline methods, the DNS model trained with the MSE-based loss \eqref{MT} proposed in \cite{strake2020DNS}, which is denoted as ``{\tt FCRN} \cite{strake2020DNS}", shows the best performance under both reverberation conditions by offering three \nth{1}-ranked metrics. Under both reverberation conditions, it also offers slightly better PESQ scores compared to the baseline from our previous work \cite{xu2021inter}, where real recordings are employed in DNS training. Our previous work denoted as ``{\tt FCRN}/{\tt PESQNet}, \cite{xu2021inter}" achieves two \nth{1}-ranked metrics and one \nth{2} rank and significantly outperforms the DNS3 baseline \cite{braun2020data} in speech quality measured by PESQ. The DNS3 baseline \cite{braun2020data} shows limited overall performance compared to the other baseline methods, which are reflected by around 0.2 and 0.24 points lower PESQ scores under the conditions without and with reverberation. This is supposedly caused by the weakest noise attenuation and dereverberation reflected by the lowest $\Delta\text{SNR}_\text{seg}$ and SRMR scores.
%%%%%%%%%%%%%%%%%%%%%%%%%%%%%%%%%%%%%%%%%%%%%%%%%%%%%%%%%%%%%%%%%%
\begin{figure}[t!]
	%\vspace*{1mm}
	\centering
	\includegraphics[width=0.48\textwidth]{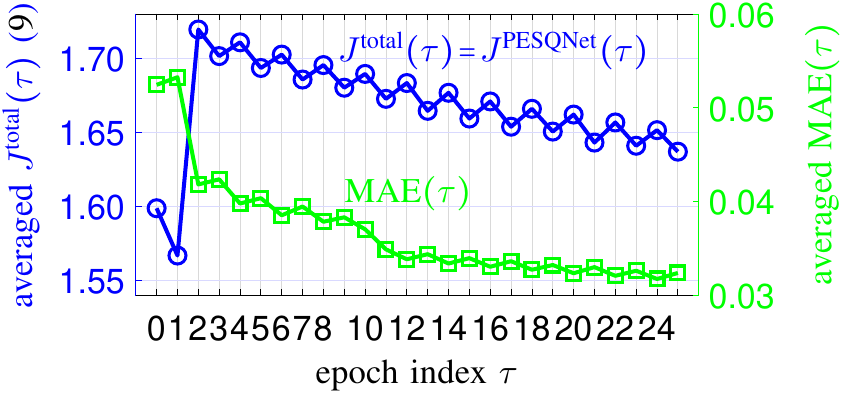}
	\caption{DNS and {\tt PESQNet} training loss performance measured on a subset of $\mathcal{D}^\text{train}_\text{DNS3}$ {\bf during \mbox{\nth{2}-stage} alternating fine-tuning.} The performance $J^\text{total}(\tau)$ \eqref{total} of the DNS is shown by the blue curve with $\alpha=0$. The performance of the {\tt PESQNet} is measured by the green curve MAE$(\tau)$.}
	\label{training_loss}
\end{figure}

For our proposed \nth{2}-stage fine-tuning employing $\alpha=0$ in \eqref{total}, our trained DNS offers the best performance among all the evaluated methods with four \nth{1}-ranked and three \nth{2}-ranked metrics. In particular, this setting can offer the best speech perceptual quality reflected by the highest PESQ scores under both reverberation conditions. Interestingly, during the \nth{2}-stage fine-tuning with our proposed $\langle1\!-\!1\rangle$ training protocol, only employing the perceptually-related loss $J_u^\text{PESQNet}$ \eqref{real} provided by our proposed {\tt PESQNet} (with $\alpha=0$ in \eqref{total}) is enough to guide the DNS in increasing the PESQ scores. Further increasing $\alpha$ towards $1$ decreases the influence of the perceptually-related loss in \eqref{total}, resulting in a gradual performance decrease of the speech quality measured by both PESQ and DNSMOS. Please note that for the hyperparameter setting $\alpha=1$, the loss functions used in the \nth{1}-stage and the \nth{2}-stage fine-tuning are exactly the same. Thus, a very similar performance is expected compared to the MSE-based reference method ``{\tt FCRN} \cite{strake2020DNS}", which is confirmed in Table\,\ref{DNS1_dev}. This setting with $\alpha=1$ serves as a ``placebo" setup intended to prove that the perceptual quality improvements for $\alpha\!<\!1$ are not just caused by the extra second-stage fine-tuning with much lower learning rate. Finally, we select the hyperparameter setting $\alpha=0$ for our proposed DNS training protocol but still report the other settings in the following.
%%%%%%%%%%%%%%%%%%%%%%%%%%%%%%%%%%%%%%%%%%%%%%%%%%%%%%
\begin{figure}[t!]
	\vspace*{-3.7mm}
	\centering
	\includegraphics[width=0.40\textwidth]{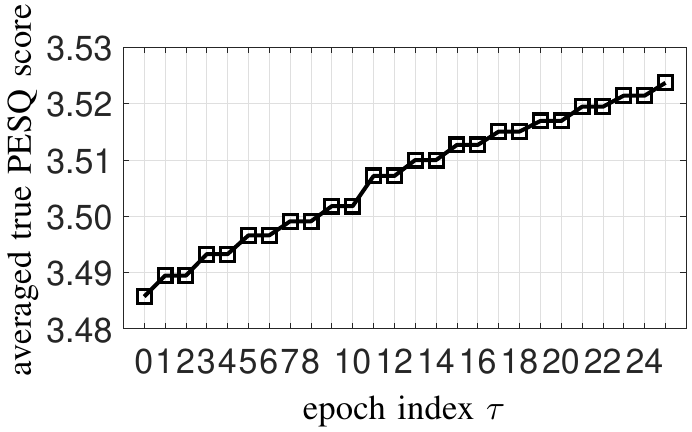}
	\caption{Averaged true ITU-T P.862.2 PESQ scores measured on a subset of $\mathcal{D}^\text{train}_\text{DNS3}$ {\bf during \mbox{\nth{2}-stage} alternating fine-tuning.} In the employed total loss \eqref{total}, $\alpha$ is set to $0$.}
	\label{training_PESQ}
\end{figure}

To illustrate the progress of the \nth{2}-stage fine-tuning for our DNS and the {\tt PESQNet}, we plot the training performance measured on a subset of the $\mathcal{D}^\text{train}_\text{DNS3}$ in Fig.\ \ref{training_loss} with the horizontal axis representing the epoch index $\tau$. The performance of the DNS is measured by the total loss $J^\text{total}=J^\text{PESQNet}$ \eqref{total}, \eqref{real}, due to $\alpha=0$, represented by the blue curve. The corresponding performance of the {\tt PESQNet} is measured by the MAE between the predicted and the ground truth PESQ scores, shown by the green curve. The initial performance of the DNS and the corresponding {\tt PESQNet} before the \nth{2}-stage alternating fine-tuning is represented by the markers at $\tau=0$. The DNS is trained with a fixed {\tt PESQNet} on epochs with odd index number, e.g., $\tau\in\left\{1,3,\ldots,25\right\}$, while the {\tt PESQNet} is trained adapting to the fixed updated DNS on epochs with even index number. Please note that an {\it inaccurately} estimated PESQ score, which is close to $\text{PESQ}_\text{max}$, can lead to a low loss value of $J^\text{total}$ \eqref{total}. Meanwhile, a more precisely estimated PESQ score obtained from the updated {\tt PESQNet} may result in an increased $J^\text{total}$. This explains the large performance variation at $\tau=2$, where the {\tt PESQNet} is trained with a significant improvement and subsequently leads to the increase of $J^\text{total}$, which then again decrease over the epochs. Most important, however, we can see that the performance of our DNS and the {\tt PESQNet} improves alternatingly during the \nth{2}-stage fine-tuning.

Furthermore, in Fig.\,\ref{training_PESQ}, we plot the averaged true PESQ scores measured on the same dataset used in Fig.\ \ref{training_loss} during the \nth{2}-stage fine-tuning. Compared to the initial PESQ performance of the DNS (marker at $\tau=0$), each epoch of the DNS training mediated by the proposed {\tt PESQNet} can achieve a better PESQ score. The PESQ improvement obtained from each learning step of DNS training is small, which is supposedly due to the small learning rate\footnote{We found that increasing the learning rate for the DNS and {\tt PESQNet} alternating training will not lead to a more significant PESQ improvement but may result in a destabilized training with a strongly fluctuating training loss. Thus, with a high learning rate, we only achieve a marginal performance improvement on the enhanced speech quality.} used in the \nth{2}-stage fine-tuning, however, it accumulates over the epochs. Please note that the DNS before the \nth{2}-stage fine-tuning is exactly the baseline ({\tt FCRN} \cite{strake2020DNS}) trained with the MSE-based loss, which offers the best performance among all the baselines, especially on PESQ scores. Accordingly, employing a small learning rate for the \nth{2}-stage fine-tuning can supposedly further increase the PESQ score without harming the performance on other metrics. Furthermore, we can infer from Fig.\,\ref{training_PESQ} that further PESQ improvement could be achieved with more training epochs for the \nth{2}-stage fine-tuning. This also proves that the perceptually related loss \eqref{real} provided by our {\tt PESQNet} can be advantageously employed for improving the perceptual quality of the DNS, and potentially of any DNS network topology.
%%%%%%%%%%%%%%%%%%%%%%%%%%%%%%%%%%%%%%%%%%%%%%%%%%%%%%%%%%%%%%%%%%
\begin{table}[t!]
	\centering
	\caption{\textbf{Instrumental quality results} on the \textbf{test set} $\mathcal{D}^\mathrm{test}_\mathrm{DNS2}$, employing \textbf{synthetic data}. DNSMOS is adopted from \cite{reddy2021dnsmos}. Best results are in {\bf bold} font, and the second best are \underline{underlined}. Proposed method with $*$.}
	\setlength\tabcolsep{2pt}
	\begin{tabular}{c c c c c}
		\hline
		& \stz{Method} & PESQ & DNSMOS  &  \stz{STOI}\\ \hline
		\multirow{4}{*}{\rotatebox{90}{REF\ }}& \multicolumn{1}{l}{Noisy} & \stz{2.37} & 3.08 & 0.88\\ \hhline{~~~~~}
		& \multicolumn{1}{l}{\stz{DNS3 Baseline \cite{braun2020data}}} & \stz{3.14} & 3.52 & 0.91\\ \hhline{~~~~~}
		& \multicolumn{1}{l}{\stz{{\tt FCRN} \cite{strake2020DNS}}} & \stz{3.25} & 3.60 & {\bf 0.93}\\ \hhline{~~~~~}
		& \multicolumn{1}{l}{\stz{{\tt FCRN}/{\tt PESQNet} \cite{xu2021inter}}} & \stz{3.24} & {\bf 3.67} & \underline{0.92}\\ \hhline{-----}
		\multirow{3}{*}{\rotatebox{90}{NEW}}& \multicolumn{1}{l}{\stz{{\tt FCRN}/{\tt PESQNet}$^*$, $\langle1\!-\!1\rangle$, $\alpha\!=\!0$}} & \stz{\bf 3.34} & \underline{3.65} & {\bf 0.93}\\ \hhline{~~~~~}
		& \multicolumn{1}{l}{\stz{{\tt FCRN}/{\tt PESQNet}, $\langle1\!-\!1\rangle$, $\alpha\!=\!0.5$}} & \stz{\underline{3.32}} & 3.63 & {\bf 0.93}\\ \hhline{~~~~~}
		& \multicolumn{1}{l}{\stz{{\tt FCRN}/{\tt PESQNet}, $\langle1\!-\!1\rangle$, $\alpha\!=\!1$}} & \stz{3.26} & 3.60 & {\bf 0.93}\\ \hline
	\end{tabular}
	\label{DNS2_test}
\end{table}
%%%%%%%%%%%%%%%%%%%%%%%%%%%%%%%%%%%%%%%%%%%%%%%%%%%%%%%%%%%
\begin{figure}[t!]
	\centering
	\includegraphics[width=0.49\textwidth]{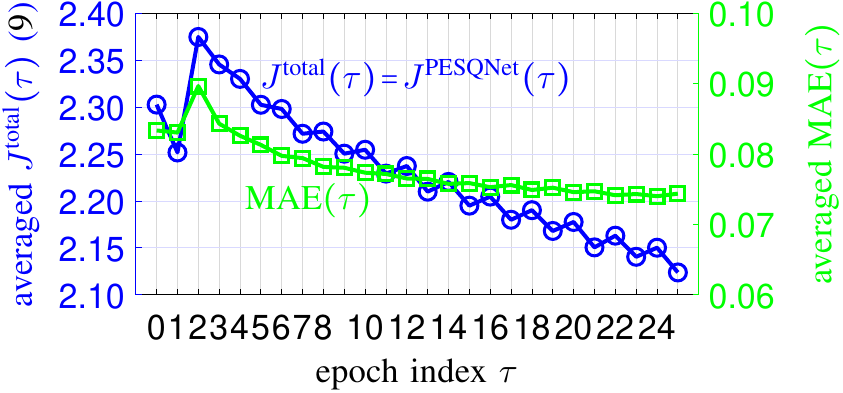}
	\caption{DNS and {\tt PESQNet} loss measured on $\mathcal{D}^\text{test}_\text{DNS2}$ {\bf during \mbox{\nth{2}-stage} alternating fine-tuning.} The performance $J^\text{total}(\tau)$ \eqref{total} of the DNS is shown by the blue curve with $\alpha=0$. The performance of the {\tt PESQNet} is measured by the green curve MAE$(\tau)$.}
	\label{test_loss}
\end{figure}
%%%%%%%%%%%%%%%%%%%%%%%%%%%%%%%%%%%%%%%%%%%%%%%%%%%%%%%%%%
\begin{figure}[t!]
	\centering
	\includegraphics[width=0.40\textwidth]{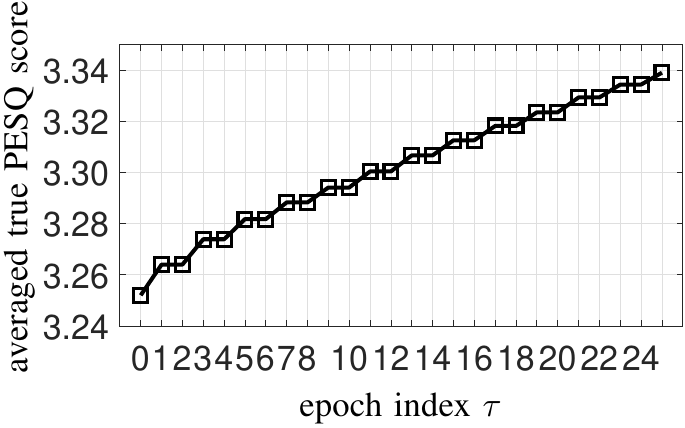}
	\caption{Averaged true ITU-T P.862.2 PESQ scores measured on $\mathcal{D}^\text{test}_\text{DNS2}$ {\bf during \mbox{\nth{2}-stage} alternating fine-tuning.} In the employed total loss \eqref{total}, $\alpha$ is set to $0$.}
	\label{test_PESQ}
\end{figure}
%%%%%%%%%%%%%%%%%%%%%%%%%%%%%%%%%%%%%%%%%%%%%%%%%%%%%%%%%%%%%%%%%%%%%%%%%%%%%%%%%%%%%%%
\subsubsection{Performance Evaluation on Synthetic Test Data}
We evaluate our \nth{2}-stage fine-tuned DNS on both synthetic and real test data. First, we look at the instrumental quality results measured on synthetic test data $\mathcal{D}^\mathrm{test}_\mathrm{DNS2}$, with results shown in Table\,\ref{DNS2_test}. The DNS fine-tuned with our novel $\langle1\!-\!1\rangle$ training protocol with $\alpha\!=\!0$ offers the best or the second-best results in all three implemented metrics. Especially, our proposed method can offer the highest PESQ score among all the evaluated methods. Compared to the baseline DNS trained with the MSE-based loss \eqref{MT} denoted as ``{\tt FCRN} \cite{strake2020DNS}", which has a PESQ score of $3.25$, we further increase PESQ by about $0.1$ points with the additional \nth{2}-stage fine-tuning employing the proposed {\tt PESQNet}. Furthermore, our \nth{2}-stage fine-tuned DNS outperforms the DNS3 baseline \cite{braun2020data} by $0.2$ PESQ points. Our proposed method also delivers the \nth{2}-ranked DNSMOS, which is only slightly lower than our previous work \cite{xu2021inter} and is $0.13$ and $0.05$ points higher than the DNS3 baseline and the ``{\tt FCRN} \cite{strake2020DNS}", respectively. 

As observed in \cite{xu2021inter}, the baseline from our previous work ({\tt FCRN}/{\tt PESQNet} \cite{xu2021inter}) cannot offer PESQ improvement on synthetic test data compared to the DNS trained with the MSE-based loss \eqref{MT} \cite{strake2020DNS} due to the instable training problem. Now we solved this issue with the novel $\langle1\!-\!1\rangle$ training protocol with gradient accumulation and significantly improve the PESQ performance compared to our previous work \cite{xu2021inter} and the same DNS trained with only the MSE-based loss \cite{strake2020DNS}.

As before, in Figs.\,\ref{test_loss} and \ref{test_PESQ}, we plot the performance of our DNS and the {\tt PESQNet} as well as the averaged true PESQ scores measured on $\mathcal{D}^\mathrm{test}_\mathrm{DNS2}$ during the \nth{2}-stage fine-tuning for $\alpha\!=\!0$. In Fig.\,\ref{test_loss}, the performance of our DNS and the proposed {\tt PESQNet} alternatingly improves during the fine-tuning, which is reflected by the decreasing $J^\text{total}(\tau)$ \eqref{total} (blue curve) and MAE$(\tau)$ (green curve), respectively. At the end of the \nth{2}-stage fine-tuning (markers at $\tau=25$), both the DNS and the {\tt PESQNet} perform better compared to their initial performance at $\tau=0$. This is also reflected in Fig.\,\ref{test_PESQ} by the averaged true PESQ scores, where each epoch of DNS fine-tuning (on epochs with odd index number) can deliver a better PESQ score. From both Figs.\,\ref{test_loss} and \ref{test_PESQ}, we can see that the progress of the \nth{2}-stage fine-tuning is very stable and smooth, which is due to the contribution of our newly proposed alternating training protocol with gradient accumulation. Furthermore, from the curves in Figs.\,\ref{test_loss} and \ref{test_PESQ}, it can be inferred that the performance of our DNS and the {\tt PESQNet} could be improved until epoch $25$, and potentially one could achieve an even higher PESQ score employing a longer \nth{2}-stage fine-tuning with more epochs.
%%%%%%%%%%%%%%%%%%%%%%%%%%%%%%%%%%%%%%%%%%%%%%%%%%%%%%%%%%%%%%%%%%
\subsubsection{Performance Evaluation on Real Test Data}
We measure the perceptual quality employing DNSMOS \cite{reddy2021dnsmos} on real test data $\mathcal{D}^\mathrm{test}_\mathrm{DNS3}$ for our \nth{2}-stage fine-tuned DNS in Table\,\ref{DNS3_dev}. On real data, the other intrusive metrics from Tables\,\ref{DNS1_dev} and \ref{DNS2_test} are not applicable. Our proposed method with $\alpha\!=\!0$ offers the best perceptual quality reflected by the highest DNSMOS score compared to all the baselines. Interestingly, compared to the baseline from our previous work \cite{xu2021inter}, which focuses on improving the performance on real recordings by using real data during training employing {\tt PESQNet}, our proposed framework trained with only synthetic data can still gain $0.02$ DNSMOS points. This is supposedly due to the contribution from the stabilized \nth{2}-stage fine-tuning, as shown in Figs.\,\ref{training_loss} and \ref{training_PESQ}, employing our novel $\langle1\!-\!1\rangle$ training protocol with gradient accumulation. Thus, from all the experimental evidence, we successfully solved the unstable training problem addressed in \cite{fu2019learning} and \cite{xu2021inter} by using our novel alternating training protocol with gradient accumulation. Furthermore, from the stabilized performance improvement on synthetic test data shown in Figs.\,\ref{test_loss} and \ref{test_PESQ}, we can also infer that the perceptual quality improvement on real test data could be more significant employing an extended \nth{2}-stage fine-tuning with more epochs.
%%%%%%%%%%%%%%%%%%%%%%%%%%%%%%%%%%%%%%%%%%%%%%%%%%%%%%%%%%%%%%%
\begin{table}[t!]
	%\large
	\centering
	\caption{\textbf{Perceptual quality measure DNSMOS \cite{reddy2021dnsmos}} on the \textbf{test set} $\mathcal{D}^\mathrm{test}_\mathrm{DNS3}$, employing \textbf{real recordings}. Best results are in {\bf bold} font, and the second best are \underline{underlined}. Proposed method with $*$}
	\setlength\tabcolsep{4pt}
	%\resizebox{0.98\linewidth}{!}{
	\begin{tabular}{c c c}
		\hline
		&Methods & \stz{DNSMOS}\\ \hhline{---}
		\multirow{4}{*}{\rotatebox{90}{REF\ }}& \multicolumn{1}{l}{Noisy}            &  \stz{2.90}  \\ \hhline{~~~}
		& \multicolumn{1}{l}{\stz{DNS3 Baseline \cite{braun2020data}}}             &  \stz{3.24}  \\ \hhline{~~~}
		& \multicolumn{1}{l}{\stz{{\tt FCRN} \cite{strake2020DNS}}}             &  \stz{3.30}  \\ \hhline{~~~}
		& \multicolumn{1}{l}{\stz{{\tt FCRN}/{\tt PESQNet}} \cite{xu2021inter}}             &  \stz{3.31}  \\ \hhline{---}
		\multirow{3}{*}{\rotatebox{90}{NEW}}	& \multicolumn{1}{l}{\stz{{\tt FCRN}/{\tt PESQNet}$^*$}, $\langle1\!-\!1\rangle$, $\alpha\!=\!0$}             &  \stz{\bf 3.33}  \\ \hhline{~~~}
		& \multicolumn{1}{l}{\stz{{\tt FCRN}/{\tt PESQNet}}, $\langle1\!-\!1\rangle$, $\alpha\!=\!0.5$}             &  \stz{\underline{3.32}}  \\ \hhline{~~~}
		& \multicolumn{1}{l}{\stz{{\tt FCRN}/{\tt PESQNet}}, $\langle1\!-\!1\rangle$, $\alpha\!=\!1$}             &  \stz{3.29}  \\ \hhline{---}
		
	\end{tabular}
	%}
	\label{DNS3_dev}
\end{table}

As expected, our fine-tuned DNS mediated by the {\tt PESQNet} ($\alpha\!=\!0$) outperforms the same DNS trained with the MSE-based loss \eqref{MT} \cite{strake2020DNS} by $0.03$ DNSMOS points. This is supposedly due to the contribution of the perceptually related loss component in the total loss \eqref{total}, which is offered by our non-intrusive {\tt PESQNet}. The perceptual quality improvement is more significant compared to the DNS3 baseline with an increase of about $0.1$ DNSMOS points on real data. Still, one could ask ``why do you {\it only} gain $0.03$ DNSMOS points w.r.t.\ the {\tt FCRN} \cite{strake2020DNS}?" The answer is: A PESQNet loss has only limited predictive power to increase a DNSMOS metric. If the DNSMOS DNN predictor would have been available to us, it could have been used in place of the {\tt PESQNet}, and a more significant DNSMOS improvement on real data would have been achieved. {\it With our proposed training framework, however, we claim to have shown a way how to effectively make use of real data in DNS training.}
%%%%%%%%%%%%%%%%%%%%%%%%%%%%%%%%%%%%%%%%%%%%%%%%%%%%%%
\section{Conclusions}
This work illustrated the benefits obtained from training a deep noise suppression (DNS) neural network mediated by an end-to-end non-intrusive {\tt PESQNet}. The employed non-intrusive {\tt PESQNet} can estimate the perceptual evaluation of speech quality (PESQ) scores of the enhanced speech signal and serves to provide a reference-free perceptual loss mediating the DNS training to maximize the PESQ score of the enhanced speech signal. We illustrate the potential of our proposed method by training a complex mask-based fully convolutional recurrent neural network ({\tt FCRN}) for the DNS task. As an important novelty, we propose to train the {\tt FCRN} and the {\tt PESQNet} alternatingly with a novel training protocol employing gradient accumulation to keep the {\tt PESQNet} up-to-date. Our proposed method is compared to several baselines, among them the same {\tt FCRN} trained with an MSE-based loss for joint denoising and dereverberation, as well as the Interspeech 2021 DNS Challenge baseline. Detailed analyses suggest that the {\tt FCRN} trained mediated by our proposed {\tt PESQNet} employing the novel alternating training protocol can further increase the PESQ performance by about 0.1 PESQ points on synthetic test data and by 0.03 DNSMOS points on real test data, both compared to training with the MSE-based loss. We excel the Interspeech 2021 DNS Challenge baseline by 0.2 PESQ points on synthetic test data and about 0.1 DNSMOS points on real test data.
\ifCLASSOPTIONcaptionsoff
  \newpage
\fi

%\clearpage
%
\bibliographystyle{IEEEtran}
\bibliography{mainTrans21}

%\end{IEEEbiography}
%% insert where needed to balance the two columns on the last page with
%%\newpage


%\vfill


\end{document}